\documentclass[10pt]{iopart}
\usepackage{graphicx}
\usepackage{color}
\usepackage{amsmath}

\newcommand{\beq}{\begin{equation}}
\newcommand{\eeq}{\end{equation}}
\newcommand{\be}{\begin{equation}}
\newcommand{\ee}{\end{equation}}
\newcommand{\beqn}{\begin{eqnarray}}
\newcommand{\eeqn}{\end{eqnarray}}
\newcommand{\bea}{\begin{eqnarray}}
\newcommand{\eea}{\end{eqnarray}}
\newcommand{\bearr}{\begin{array}}
\newcommand{\enarr}{\end{array}}

\newcommand{\eps}{\varepsilon}

\begin{document}
\paper[]{Numerical analysis of heat transport in classical one-dimensional systems}

\date{\today}

\author{Antonio Politi$^{1,2}$}

\address{$^1$Istituto dei Sistemi Complessi, Consiglio Nazionale
delle Ricerche, via Madonna del Piano 10, I-50019 Sesto Fiorentino, Italy}

\address{$^2$Institute for Complex Systems and Mathematical Biology
           University of Aberdeen, Aberdeen AB24 3UE, United Kingdom}

\ead{a.politi@abdn.ac.uk}

\begin{abstract}
Numerical studies of some unidimensional systems suggest 
that Fourier law is satisfied, where theory predicts a divergence 
of heat conductivity with the system size.
Here, I revisit some such models, finding that in all cases
a divergence asymptotically emerges.
This includes a variant of the ding-a-ling model, where I find that, contrary
to previous claims, the ``anomalous" growth starts already 
for moderate system sizes. 
More conceptually interesting is the case of non-binding potentials,
whose behavior is well reproduced by assuming that the energy flux across the 
nonequilibrium stationary state is the sum of two contributions: a diffusive and
a hydrodynamic one.
This approach, which extends an idea previously formulated for nearly
integrable systems, allows to conclude that the asymptotic regime is always
dominated by the anomalous hydrodynamic component, but the crossover may occur
for extremely long system sizes.

\end{abstract}

\maketitle

\section{Introduction}

Since numerical evidence of anomalous heat conductivity was first given in chains of classical nonlinear oscillators~\cite{LLP97},
much progress has been made both by simulating nonequilibrium steady states and by studying fluctuations of equilibrium
regimes~\cite{Lepri03,Wang2011,Liu2014}.
On the theoretical side, the key cornerstones are based on mode coupling theories~\cite{Narayan02,Delfini06}, 
kinetic approaches~\cite{Pereverzev2003,Nickel07,Lukkarinen2008}, and fluctuating-hydrodynamic equations~\cite{Spohn2014}.

The typical setup where heat conduction is investigated consists of a chain of oscillators interacting via 
nearest-neighbour forces (mediated by a potential $V(u)$, where
$u$ is the mutual distance), which exchange energy with two thermal reservoirs
located at its extrema and operating at different temperatures.
It is well established that whenever 
energy is the only conserved quantity, normal diffusion and normal conductivity are expected: i.e. the energy
flux is inversely proportional to the system size $L$.
A chain of oscillators interacting with an external substrate is a typical example thereof.

The presence of a second conservation law, namely linear momentum (a consequence of translational invariance),
may induce a diverging, anomalous, heat conductivity. However, this divergence can arise only if
a further conservation law is present: the so called total stretch, defined as the sum of the mutual particle
separations $u$.
In practice, this is automatically ensured when $u$ is a physical length (either longitudianal or
transversal). Normal conductivity is, instead, observed when $u$ is an angular variable, as for instance in coupled 
rotors~\cite{Giardina99,Gendelman2000}.
In this case, the total stretch being the sum of all mutual angle separations can diffuse, since a variation of $2\pi$ 
does not have any physical implication on the dynamics.
The overall scenario is well captured by the fluctuating-hydrodynamics approach~\cite{Spohn2014}
based on three coupled fields, associated to three conservation laws (energy, linear momentum, and total stretch).

According to theory, thermal conductivity generically diverges as $L^{1/3}$.
The major exception is represented by systems with symmetric potential (and zero pressure);
the most renown example is the FPUT-$\beta$ model, where $V(u) = u^2/2+u^4/4$. 
In such a subclass, fluctuating hydrodynamics predicts an $L^{1/2}$ divergence, while kinetic theories predict an exponent 
2/5~\cite{Pereverzev2003,Dematteis2020}. 
Numerical simulations confirm the latter prediction~\cite{Delfini08c,Takatsu2024}.

Finally, there exists the special class of integrable systems; they are typically characterized by
a ballistic dynamics, where the flux is independent of the system size, the reason
being that the independent modes travel undamped from one to the other boundary of the chain.
However, integrable systems with symmetric potentials may again display an unexpected behavior.
This is, for instance, the case of the hard-point chain, where the potential $V(u)$ is an infinitely
high box of width $w$. If the particles have all the same mass, each collision, occurring either when $u=0$, or $u=w$ 
amounts to an exchange of velocities.
Hence, the dynamics is characterized by conserved quantities: the ``velocitons" which jump from one to another
particle, when a collision occurs.
If the average interparticle distance is set equal to $w/2$, the velocitons do not move in a preferential direction;
they diffuse in an anomalous way, leading to a non ballistic divergence of the heat conductivity~\cite{Politi2011}.

The general scenario has been confirmed by several numerical simulations, but some results are at odds
with the theoretical predictions.
Here, I focus on the typical inconsistency: the evidence of a seemingly finite conductivity in cases
where theory predicts a divergence (see e.g.,~\cite{Spohn2014}).
I show that many numerical observations are affected by strong finite-size effects,
which mask the asymptotic divergence. In practice, in many models, the effective heat conductivity
is the sum of two contributions, a kinetic one sustained by phonon-like modes characterized by a
long mean-free path (MFP). This idea was already proposed in~\cite{Lepri2020} to explain heat transport
in nearly integrable systems. Therein, it is natural to expect phonon-like modes to
propagate over long distances before dissipating their energy. Here, I show that the same decomposition
applies to a much wider class of systems, namely models characterized by non-binding energy which
are not nearly integrable, as revealed by their fully chaotic spectrum of Lyapunov exponents.

In Sec.~\ref{sec:methods}, the relevant observables are introduced, together with the mathematical and numerical tools.
In Sec.~\ref{sec:quasi}, I briefly revisit the conjecture introduced in Ref.~\cite{Lepri2020} to describe
the behavior of nearly integrable systems, rephrasing it, to introduce its application in the following section.
In Sec.~\ref{sec:nonbinding}, the behavior of two models of oscillators interacting via non binding forces, is
analysed, showing that the effective conductivity is, to a large extent, the sum of
a normal and an diverging component.
Sec.~\ref{sec:canada} is devoted to the study of a variant of the ding-a-ling model~\cite{Casati84},
which conserves linear momentum and is often quoted as a clear example of normal conductivity~\cite{Lee-Dadswell2010}.
Contrary to the initial claims, I find compelling evidence of a diverging conductivity, in full agreement
with the theoretical expectation $L^{1/3}$. 
Finally, in Sec.~\ref{sec:conclusion}, some conclusions are drawn and two open problems briefly mentioned.

\section{Methods and tools}\label{sec:methods}

The typical model considered in this paper is ruled by the Hamiltonian
\begin{equation}
H = \sum_{n=1}^L \left[\frac{p_n^2}{2m_n} + V(q_{n+1} - q_n) \right] \, ,
\label{Hamil}
\end{equation}
where $m_n$, $q_n$ and $p_n$ represent respectively the mass, displacement and momentum of the $n$th particle. 
$L$ is used both to denote the number of particles as well as the system size (an average unit
interparticle distance is implicitly assumed).
Everywhere, I work with fixed boundary conditions.

The first (last) particle of the chain is attached to a heat bath operating
at temperature $T_L$ ($T_R$).  All of the work is based on
the study of the resulting NonEquilibrium Stationary State (NESS) for different chain lengths and, occasionally, different
parameter values.

One reason I prefer this approach with respect to the analysis of equilibrium fluctuations is that one can straightforwardly§
identify the appropriate (long) time scale, once a specific system size is given.
In fact, given a chain length $L$, it is conceptually easy to determine the required simulation time on the basis of
the corresponding statistical fluctuations of the energy flux $J$. 
A second justification, is that the flux itself can be easily computed from the energy 
exchanged by the chain with the heat baths, without the need of engaging ourselves with an appropriate bulk definition
which can be relatively tricky if the coupling extends beyond nearest neighbours
as for the model discussed in Sec.~\ref{sec:canada}.

As for the coupling with the heat baths, it is typically implemented by randomly replacing the velocity of the first and last particle
with a zero-average Gaussian-distributed variable whose standard deviation is fixed in accordance to the prescribed temperature.
This way, one can use a symplectic algorithm to integrate the equations of motion. The time separation between consecutive
thermstat actions quantifies the interaction strength. As already discussed in~\cite{Lepri03}, weak and too strong interactions
reduce the flux, making the numerical estimate less statistically reliable. In this paper,
time separation is everywhere chosen to be approximately 1 time unit. 
Moreover, for numerical convenience, it is always constant (I have verified 
that this choice does not lead to appreciable differences). 

A serious obstacle for an accurate  estimate of the average flux in large chains
is the selection of the initial condition. If it is not sampled according to the (unkwnon) stationary
distribution, one must discard a transient to allow the distribution to converge to
its asymptotic stationary shape. The larger is $L$ and the longer is this time.
Here, I adopt a trick which allows shortening this time as much as possible.
The overall analysis of a given model is performed by progressively doubling the chain length so that one can use
the outcome of the simulations for size $L$, to generate a proper initial condition for the size $2L$.
More precisely, let $\{q_i^{(a)},p_i^{(a)}\}_L$ and $\{q_i^{(b)},p_i^{(b)}\}_L$ be two NESS configurations generated
while studying a chain of length $L$ well separated in time to be considered statistically independent.
The initial condition
$\{ \overline{q}_i,\overline{p}_i\}_{2L}$ of the chain of length $2L$ is built by alternating the momenta 
of configuration $(a)$ with those of configuration $(b)$. This way, the resulting temperature profile is as
close as possible to the expected asymptotic one.
As for the positions, the procedure is a bit more involved, since the potential energy depends on the
relative distances. In practice, I first determine $u_i^{(a)}=q_i^{(a)}-q_{i-1}^{(a)}$ and $u_i^{(b)}=q_i^{(b)}-q_{i-1}^{(b)}$
to then obtain $\overline{u}_i$ by alternating the two $u_i$ sequences and finally generate $\overline{q}_i$ by integrating
in space the $\overline{u}_i$ separations. A final micro adjustment is made to the $\overline{q}_i$ variables to ensure
the desired chain length.
This way, the initial configuration is a good approximation of a generic stationary state and, in fact, simulations do not
reveal any appreciable transient in the simulation of the longer chain.

Finally, by denoting with $J(L)$ the heat flux in a system of size $L$ and with $\Delta T=T_L-T_R$ the temperature difference ($T_L$ is
assumed to be larger than $T_R$), the effective conductivity can be defined as
\be
\kappa(L) = \frac{J(L) L}{\Delta T}
\ee
since $\Delta T/L$ is, in fact the temperature gradient imposed by the thermostats.
A system is said to satify Fourier law if $\kappa(L)$ is independent of $L$ for $L\to\infty$.
If, instead, $\kappa \approx L^\alpha$, the system is said to exhibit anomalous conductivity.
As anticipated in the introduction, in the widest class of anomalous systems $\alpha =1/3$.

\section{Pseudo-normal conductivity}\label{sec:quasi}

Nearly integrable systems represent a class of models which, surprisingly, seem to 
exhibit normal conductivity~\cite{Iacobucci2010,Chen2014,Zhao2018}.
An explanation of this anomaly was given in Ref.~\cite{Lepri2020}, by arguing that in such systems, the heat resistance 
can be approximately expressed as the sum of two contributions:
(i) a kinetic term arising from the propagation of phonon-like modes; (ii) a hydrodynamic term, responsible for the 
asymptotic divergence of the conductivity.
Here, I review the method, underlining the relevant conceptual assumptions.

The simplest description of the kinetic channel is, in terms of the effective resistance,
\be
R_{kin} = R_0\left(1 + \frac{ L}{\it l}\right) \; .
\label{eq:diffuse}
\ee
Here $R_0$ can be interpreted as a contact resistance, due to the impedence mismatch between
the thermal bath and the physical system (see also~\cite{Aoki01});
$\it l$ is, instead, the MFP of the phonons.
For $L \ll{\it l}$, the behavior is ballistic, i.e. the resistance and hence the flux is independent of $L$.
For $L\gg{\it l}$, the system operates in the diffusive regime, characterized by a resistance proportional
to the system length.

The anomalous component, whenever present, consists of a resistance which grows sublinearly with $L$, i.e. 
\be
R_{an} = \frac{L^{1-\alpha}}{\sigma} \; , 
\label{eq:diffuse2}
\ee
where $\alpha>0$ identifies the universality class, while $\sigma$ gauges the overall amplitude of this term.

The core of the conjecture is that for finite lengths, the two channels of heat transport operate in parallel,
so that the effective conductivity $\kappa(L)$ is the sum of the two contributions.
More precisely, in the limit of small temperature differences, 
\be
\kappa(L) =  \frac{1}{(\Delta T)L} \left[ \frac{1}{R_{kin}} + \frac{1}{R_{an}}    \right ] =
\frac{1}{\Delta T} \left [  \frac{1/R_0}{1/L + 1/{\it l}} + \sigma L^{\alpha} \right ]
\label{eq:scalingk}
\ee
The first term in square brackets saturates at ${\it l}/R_0$, so that the
anomalous component prevails only when the length is larger than the crossover value
\be
L_c = \left(\frac{\it l}{\sigma R_0} \right)^{1/\alpha} 
\label{eq:mfp}
\ee
Depending on the various parameter values, $L_c$ may be so large to make the 
detection of the asymptotic divergence practically impossible. 
Nearly integrable systems represent a prominent instance where this assumption has been validated,
since therein the MFP $\it l$ can be tuned by adjusting the closeness to integrability.
More precisely, quantitative studies have been performed in Ref.~\cite{Lepri2020}, by analysing the 
diatomic Hard Point Gas (HPG), a system of point particles, which interact only via elastic collisions.
In that model, when the masses are all equal, the dynamics is integrable and energy is ballistically transported 
across the chain; a small mass heterogeneity induces a weak energy ``dissipation", and hence a finite
MFP.

\section{Non binding potentials}
\label{sec:nonbinding}

In this section, I show that the approach described above applies to a wider class of dynamical systems,
far from the integrable limit as testified by their clearly chaotic spectrum of Lyapunov exponents.
I refer to systems characterized by a non-binding potentials. 
Particularly intriguing are the Lennard-Jones, Morse, and Coulomb potentials~\cite{Savin2014,Gendelman2014},
because of the physical relevance of these models.

I first investigate a softened version of the HPG
(SPG: Soft Point Gas), characterized by the repulsive potential 
\be
V(u) = \frac{C}{u^2}
\label{eq:soft_pot}
\ee
where $u$ is the interparticle distance and $C$ has been set equal to 1/2 for easiness of simulations. 
The time step has been fixed to $\delta t = 0.002$ in order to reproduce the collision dynamics with sufficient accuracy.

\begin{figure}[ht!]
\centering
\includegraphics[width=0.6\textwidth,clip]{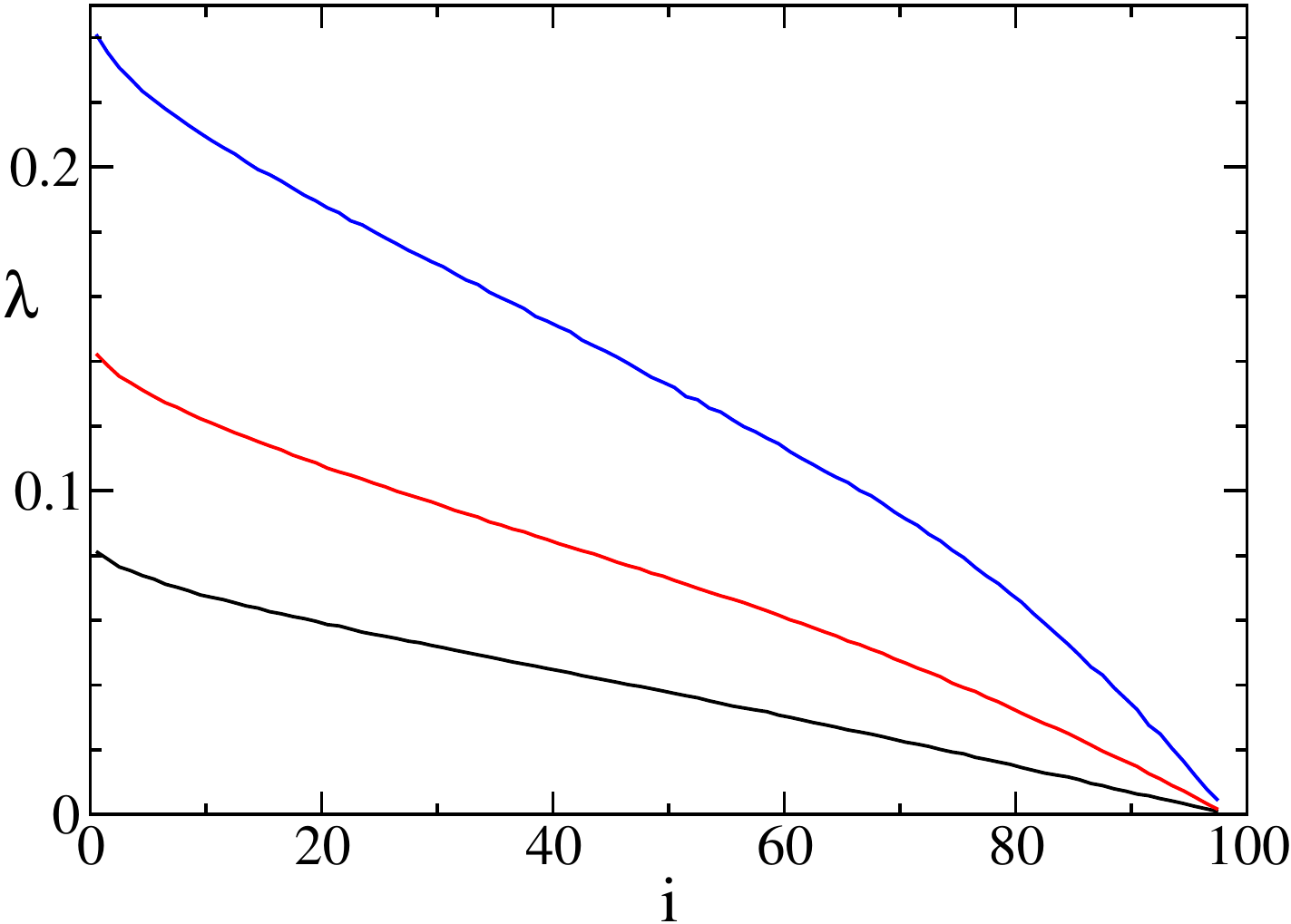}
\caption{Lyapunov spectra of the SPG for three different energy densities:
0.5, 1, 2, from bottom to top. The chain length is $L=100$ and only the first half of the spectrum is displayed}
\label{fig:lyap}
\end{figure}

At variance with the HPG, a homogeneous chain is chaotic (hence non integrable), as testified by 
the spectrum of Lyapunov exponents: see Fig.~\ref{fig:lyap}, where the first half of the spectrum
is displayed for three different energy densities.
The Lyapunov exponents are clearly larger than 0 and the overall shape is akin to that
of standard chaotic Hamiltonian systems.

\begin{figure}[ht!]
\centering
\includegraphics[width=0.6\textwidth,clip]{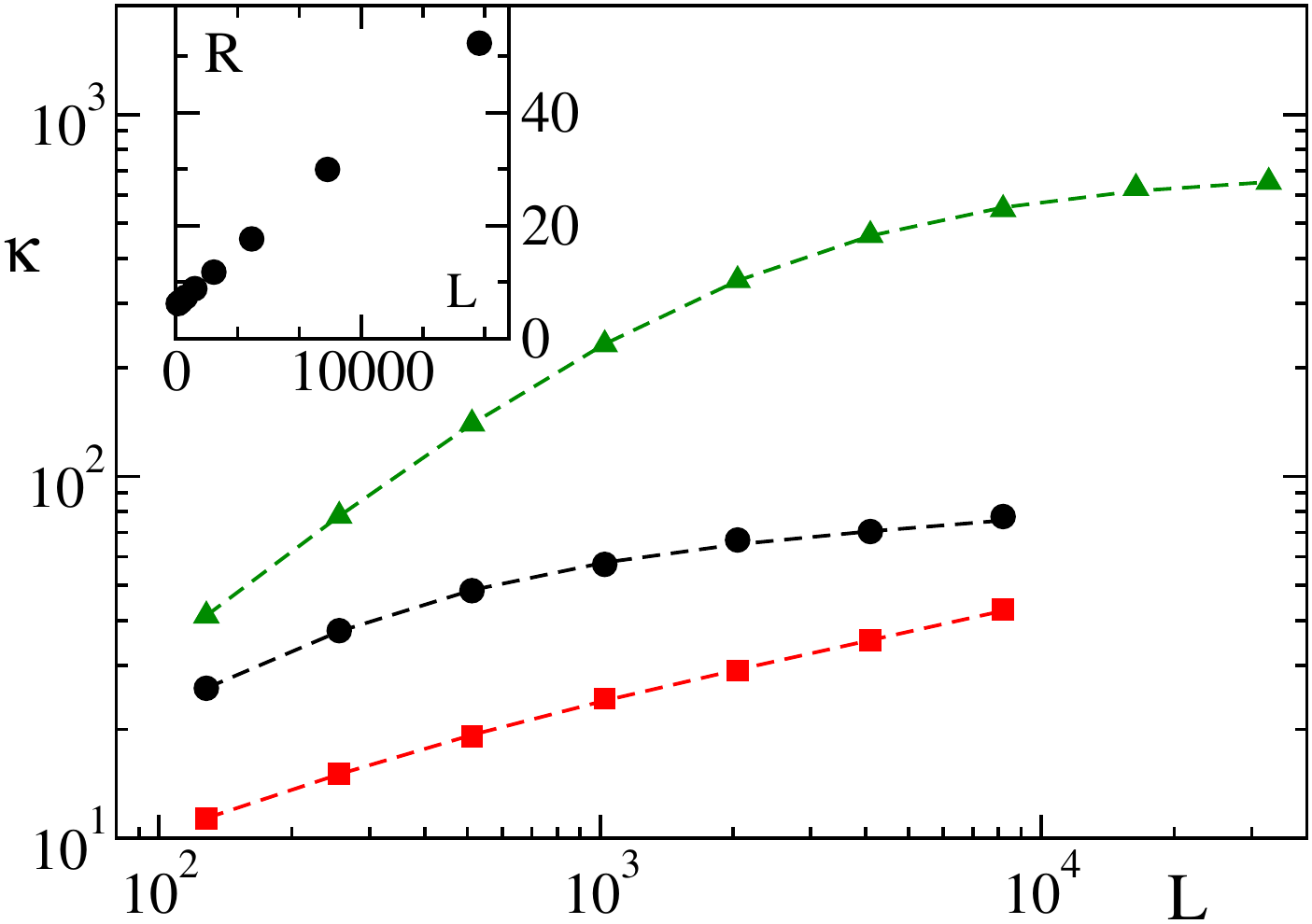}
\caption{Effective conductivity of the SPG versus system size for three
different $\eps$-values: 0, 0.1, and 0.3 from top to bottom.
Symbols refer to direct simulations, while the dashed lines are the results of a fit
with the expression (2), where $R_0$, $\it l$, and $\sigma$ are free parameters
(see Fig.~\ref{fig:meanpath} for the values of $\it l$ and $\sigma$; $R_0 \approx 6$ in the whole
range of $\eps$ values).  In the inset,
the resistance (essentially the inverse of the flux) is reported in the homogeneous
case $\eps=0$.}
\label{fig:softHPG}
\end{figure}

In spite of the system being chaotic, simulations performed by setting the
heat-bath temperatures at $T_L=1$, and $T_R=0.5$, suggest evidence of a
normal conduction, analogously to what observed in Refs.~\cite{Savin2014,Gendelman2014}.
In fact, the effective conductivity displayed in Fig.~\ref{fig:softHPG} (see the triangles)
suggests a slow saturation. The evidence is even more compelling in the inset, where the
actual resistance, plotted for various system sizes, displays a clean linear growth.

Hence, I have decided to introduce a small asimmetry between the mass of even 
($1+\eps$) and odd ($1-\eps$) particles.  In fact, as seen in \cite{Lepri2020} with reference to the HPG, 
a small $\eps$ is able to downgrade the ballistic transport, allowing for an easier detection 
of the diverging hydrodynamic component.

The resulting effective conductivities for $\eps=0.1$, and $\eps=0.3$ are again reported in Fig.~\ref{fig:softHPG}
(see dots and squares, respectively). One can see more evidence of a grwoth, but the asymptotic behavior is not clear.
It is thus tempting to use Eq.~(\ref{eq:scalingk}) to characterize the effective conductivity 
for the different degrees of mass inhomogeneity.
The dashed lines shown in Fig.~\ref{fig:softHPG} correspond to the resulting
fits, made by considering $R_0$, $\it l$, and $\sigma$ as free parameters
and setting a priori $\alpha=1/3$ ($\Delta T=0.5$ is known a priori)
The agreement is remarkably good in all cases, confirming
that the theoretical expression (\ref{eq:scalingk}) captures the crossover towards the expected
divergence, even when the phenomenon itself is not obviously visible 
in the range of lengths accessible in the simulations.

\begin{figure}[ht!]
\centering
\includegraphics[width=0.6\textwidth,clip]{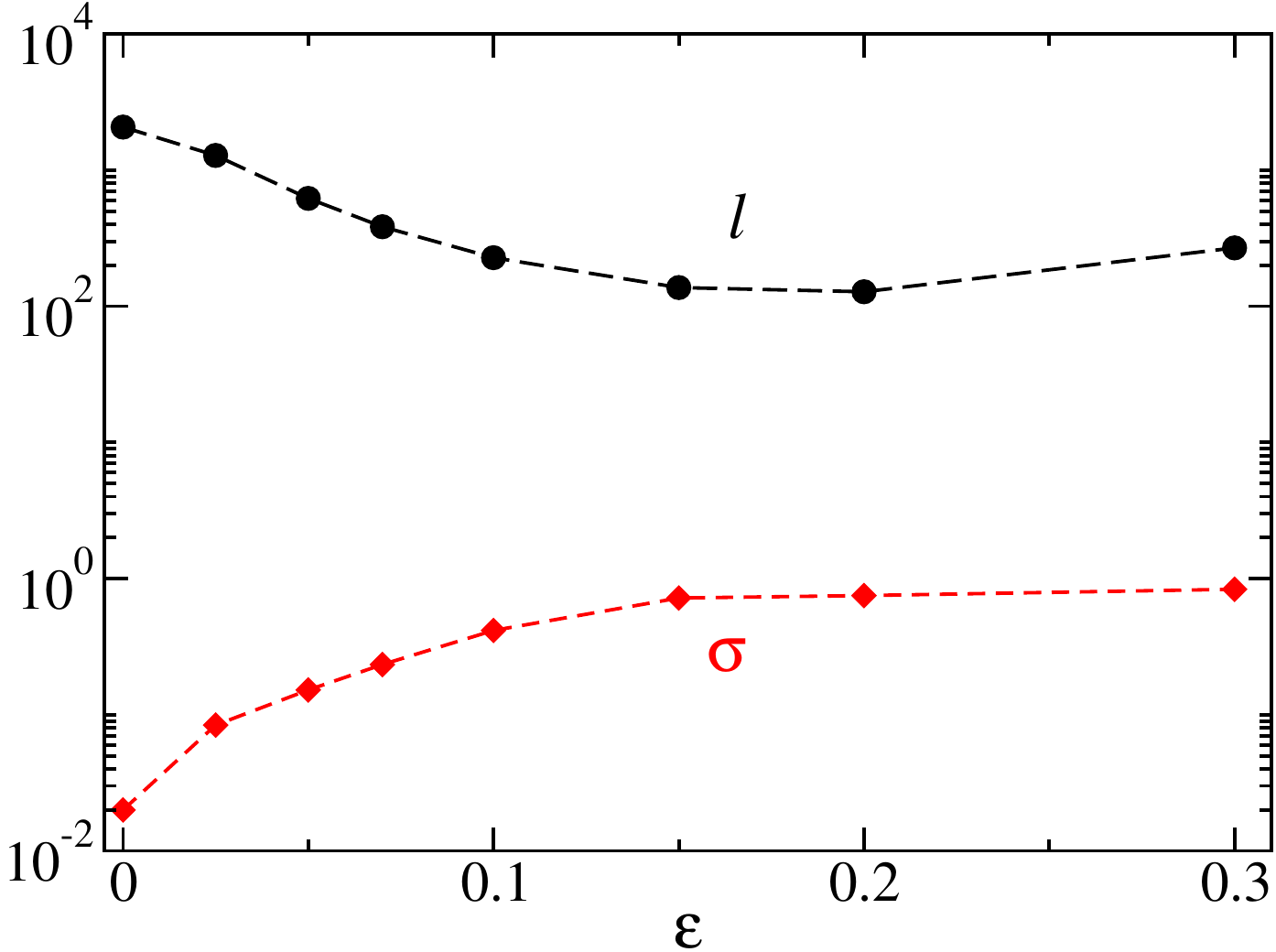}
\caption{The mean free path $\it l$ (black) and the amplitude $\sigma$ (red)
of the nonlinear term in the SPG for different degrees of mass
inhomogeneity.}
\label{fig:meanpath}
\end{figure}

The fitted parameters can be then used to investigate the dependence of the 
transport properties on the mass asymmetry $\eps$. 
The most instructive observable is the MFP:
from Fig.~\ref{fig:meanpath} (see the upper black curve), one can see that
$\it l$ grows significantly while approaching the homogeneous regime, 
without nevertheless displaying a divergence as in an integrable limit.

Further interesting information is contained in the dependence of $\sigma$ on $\eps$.
In Fig.~\ref{fig:meanpath}, one can see that $\sigma$ becomes
smaller and smaller , when $\eps \to 0$.
This contributes to further increase the crossover length $L_c$ as from Eq.~(\ref{eq:mfp}) 
($R_0$ remains steadily close to 6 when $\eps\to 0$). This is at variance with
the diatomic HPG studied in~\cite{Lepri2020}, where it is understood
that $\sigma$ does not exhibit any particular (pseudo)-singularity in
the integrable limit.

In conclusion, one can conclude that the homogeneous SPG ($\eps=0)$ is characterized by a diverging
conductivity, but a direct measure is prohibitive, since the crossover length (i.e. the length where
the anomalous component starts being comparable with the kinetic contribution) is about $2\ 10^4$.

\begin{figure}[ht!]
\centering
\includegraphics[width=0.6\textwidth,clip]{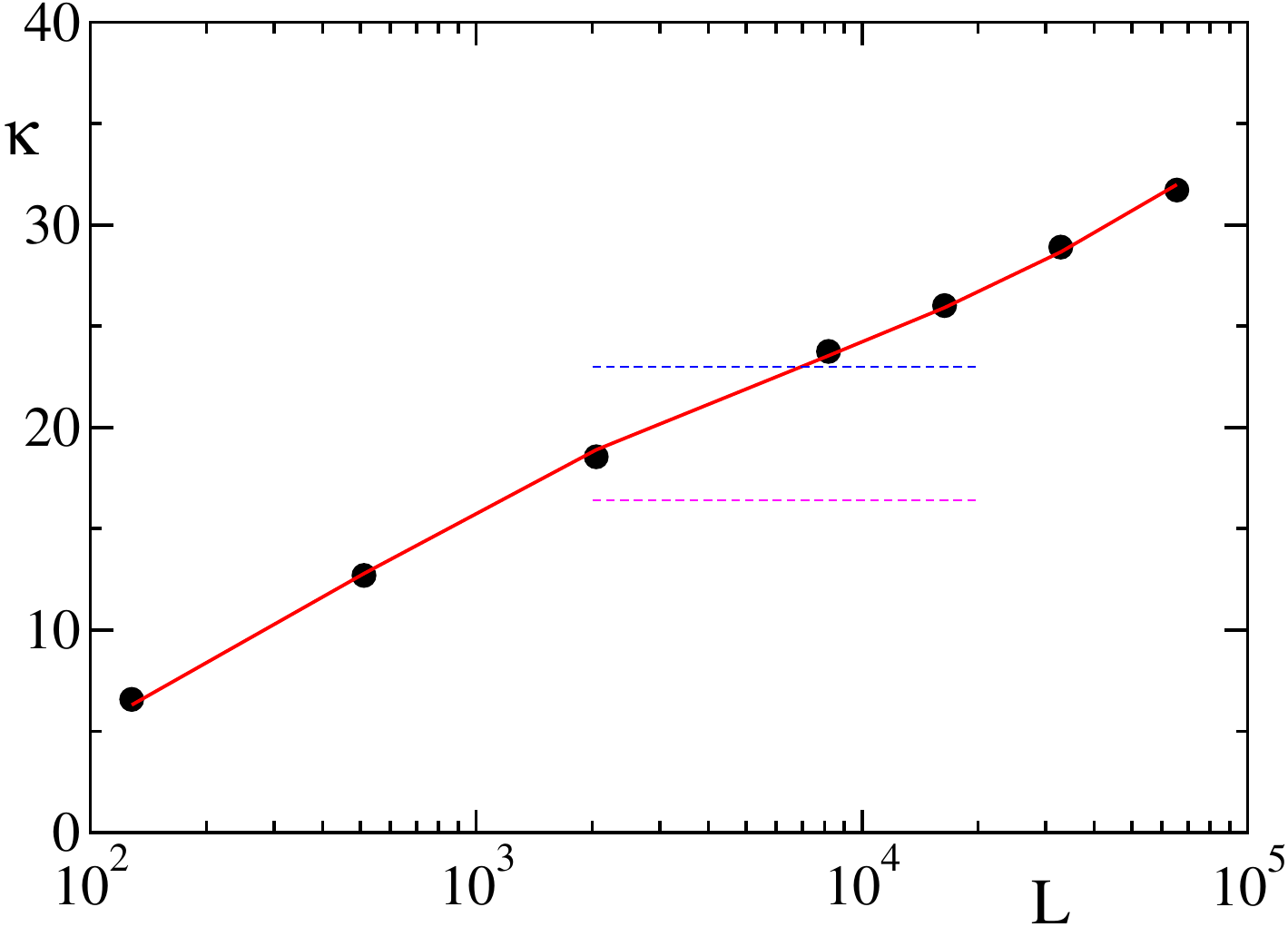}
\caption{Effective conductivity versus the chain length $L$. Full dots refer to
the outcome of direct simulations. The solid red line is the outcome of a fit
with expression 
(\ref{eq:scalingk}). 
The upper horizontal dashed line corresponds to the conductivity
determined in \cite{Gendelman2014} with kinetic arguments. The lower horizonatal line corresponds to
the conductvity extracted from the above mentioned fit.}
\label{fig:gendel}
\end{figure}

In order to strengthen the conclusion that the anomalous divergence is an ubiquituous feature within 
non-binding potentials, I have also revisited the  model thoroughly investigated in \cite{Gendelman2014}, 
where the authors have also developed kinetic arguments to estimate the presumed finite conductivity. 
The model is characterized by a repulsive potential: parabolic on one side and flat on the other,
\be
    {V(u) = }
\begin{cases}
    (u-d)^2/2 & u < d \\
    0  &  u \ge d
\end{cases}
\ee
I investigated the case $d=1$, because as claimed in \cite{Gendelman2014}, the conductivity should be independent of the temperature:
Again, the boundary temperatures are  $T_L=1$ and $T_R=0.5$.
The results are plotted in Fig.~\ref{fig:gendel}, by using the same format as in Ref.~\cite{Gendelman2014} (i.e. the vertical scale is
linear).
Up to $L=8192$, the values agree with those in Fig.~3 of Ref.~\cite{Gendelman2014} ($L=8192$ was the maximum value reported therein).
Above $L=8192$, though slowly, the effective conductivity keeps increasing.
A fit with Eq.~(\ref{eq:scalingk}) produces the red curve, which follows very closely the numerical points.

Once again, it appears that the anomalous divergence is expected to arise in this model as well.
On a quantitative level, I now compare the estimate of the (finite) conductivity presented 
in Ref.~\cite{Gendelman2014} with the diffusion component ${\it l}/R_0$, estimated from the fit. 
The two values correspond
to the two horizontal lines reported in Fig.~\ref{fig:gendel}. There is a certain disagreement,
but in spite of the very different approaches adopted to obtain them, I would conclude that they are
relatively consistent. In fact, one should note that the numerically estimated diffusive conductivity
is based on the assumption that Eq.~(\ref{eq:scalingk}) captures exactly the ``short"-length beahvior,
but it is well possible that the formula (\ref{eq:scalingk}) is too simple to reproduce perfectly well the flux
over such scales.

\section{Ding-a-ling type model}
\label{sec:canada}
Besides the class of models with non binding potentials, there exists an awkward result on a variant
of the ding-a-ling model~\cite{Casati84}.
In the ding-a-ling model, two types of particles alternate: even particles are harmonically attached to an
external substrate, while odd particles interact with the even ones via elastic collisions.
The original model exhibits a finite conductivity and this is well understood, since linear momentum is not conserved.
In Ref.~\cite{Lee-Dadswell2010}, a variant was proposed, where each even particle harmonically interacts
with the previous and following even particle.
Being all interactions internal, linear momentum is conserved and one would now expect a divergence of the
heat conductivity. Unexpectedly, the authors of \cite{Lee-Dadswell2010} found that Fourier law is satisfied.

A few years later, a variant of the the same model, without hard collisions and a different coupling scheme
was proposed and studied, finding evidence of a diverging conductivity~\cite{Gao2016}.
At a first glance, the only possible justification for the strange behavior observed in Ref.~\cite{Lee-Dadswell2010} 
is that, at variance with all other simple models, it includes next-to-nearest neighbour interactions,
but this does not seem to be a convincing explanation, since the asymptotic behavior is the natural 
result of a coarse-graining process.

I have decided to revisit the model, since it is considered a clear counterexample to the theoretical arguments.
Most of the simulations in Ref.~\cite{Lee-Dadswell2010} have been  performed in an equilibrium setup (by studying fluctuations), 
and the flux is estimated by implementing 
a ``non-standard" macroscopic formula expressed in terms of Fourier modes.
Details of the simulations are scarce, especially about the realization of the NESS.
As discussed in Sec.~\ref{sec:methods}, here I focus exclusively on nonequilibrium stationary 
simulations for various reasons,
including the opportunity to determine the energy flux in a straightforward way.

Additional details: the chain length is fixed so that the average equilibrium interparticle distance is 1.
The number $L$ of particles is odd: particle 1 ($L$) may collide with a hard wall in 0 ($L+1$) in which case
its velocity is randomly reset according to the temperature $T_L$ ($T_R$) (the velocity distribution is 
$P(v) \propto v \exp[-v^2/(2T)]$ to ensure the right sign).
The first and $L$th particle are also attached to the 2nd and $(N-1)$st particle respectively, via harmonic springs.
Finally, because of the more complex spatial structure, in this case I have not implemented the trick discussed in Sec.~\ref{sec:methods} 
to generate optimal initial conditions of length $2L$, starting from final conditions obtained for length $L$: I limited myself
to discard a sufficiently long transient.

Simulations have been performed by implementing a simple but approximate scheme: equations are integrated with a symplectic algorithm
(bilateral leap-frog). At the end of the each time step, pairs of particles which have crossed each other (because of the finite time step)
are rearranged: (i) their velocities are exchanged (to simulate the collision); the position of the free particle is shifted 
to that of the bound particle, to restore the spatial order.
This way, the total linear momentum and the total energy are perfectly conserved (as well as the total stretch).
Most of the simulations have been performed by setting the time step equal to $\delta t = 10^{-2}$, but some have been repeated for 
$\delta t=10^{-3}$, confirming the numerical results.

\begin{figure}[ht!]
\centering
\includegraphics[width=0.6\textwidth,clip]{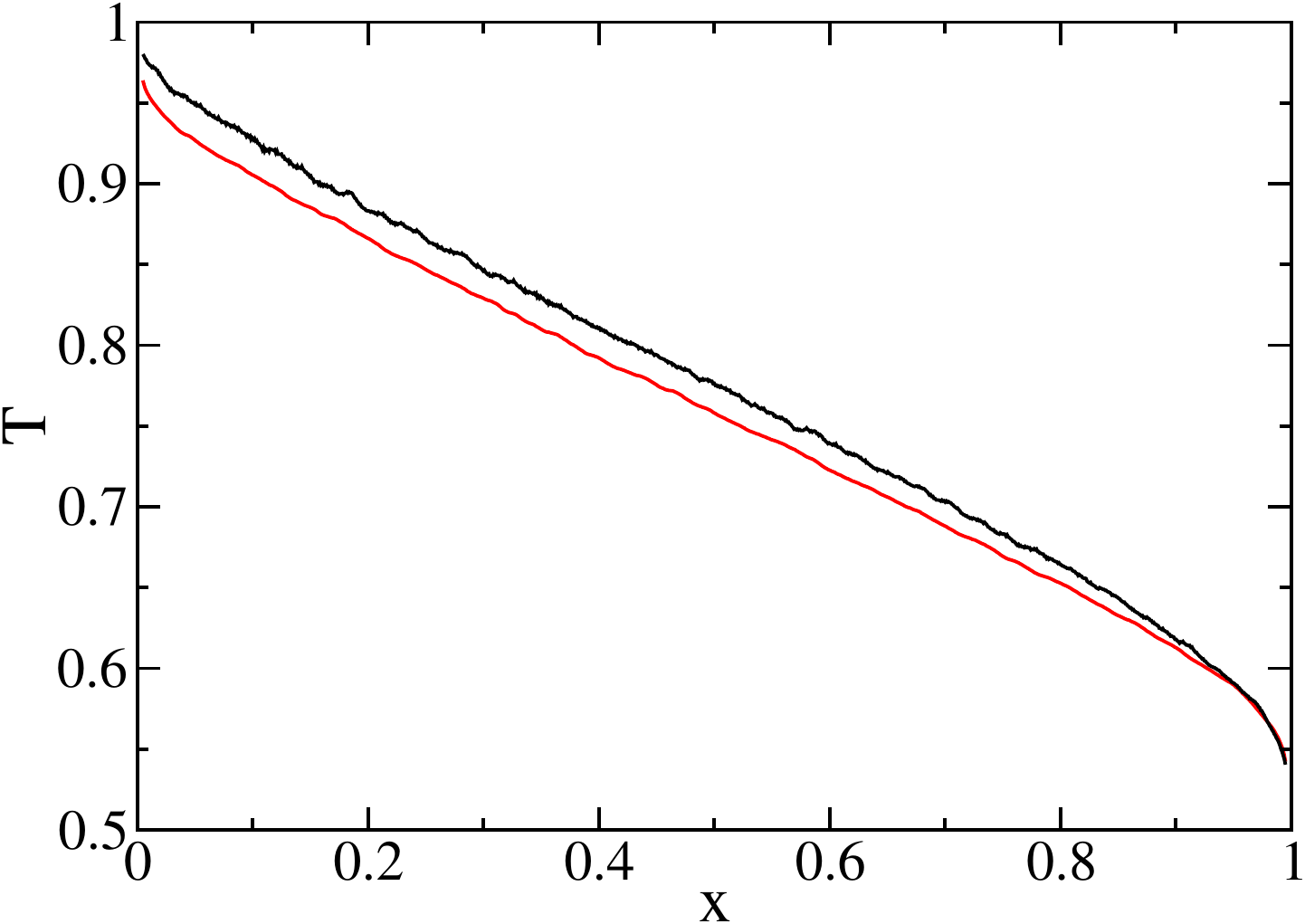}
\caption{Temperature profile for the hard (black) and soft (red) version of the model in \cite{Lee-Dadswell2010} 
versus the rescaled length $x/L$ (the chain length is 1025).}
\label{fig:canad}
\end{figure}

The outcome is shown in Fig.~\ref{fig:canad}, where the temperature profile is presented for a numerical experiment
where the heat baths operate at the temperatures $T_L=1$ and $T_R=0.5$, respectively (see the black curve)
The profile is typical of those emerging in the presence of anomalous (diverging) conductivity.

\begin{figure}[ht!]
\centering
\includegraphics[width=0.6\textwidth,clip]{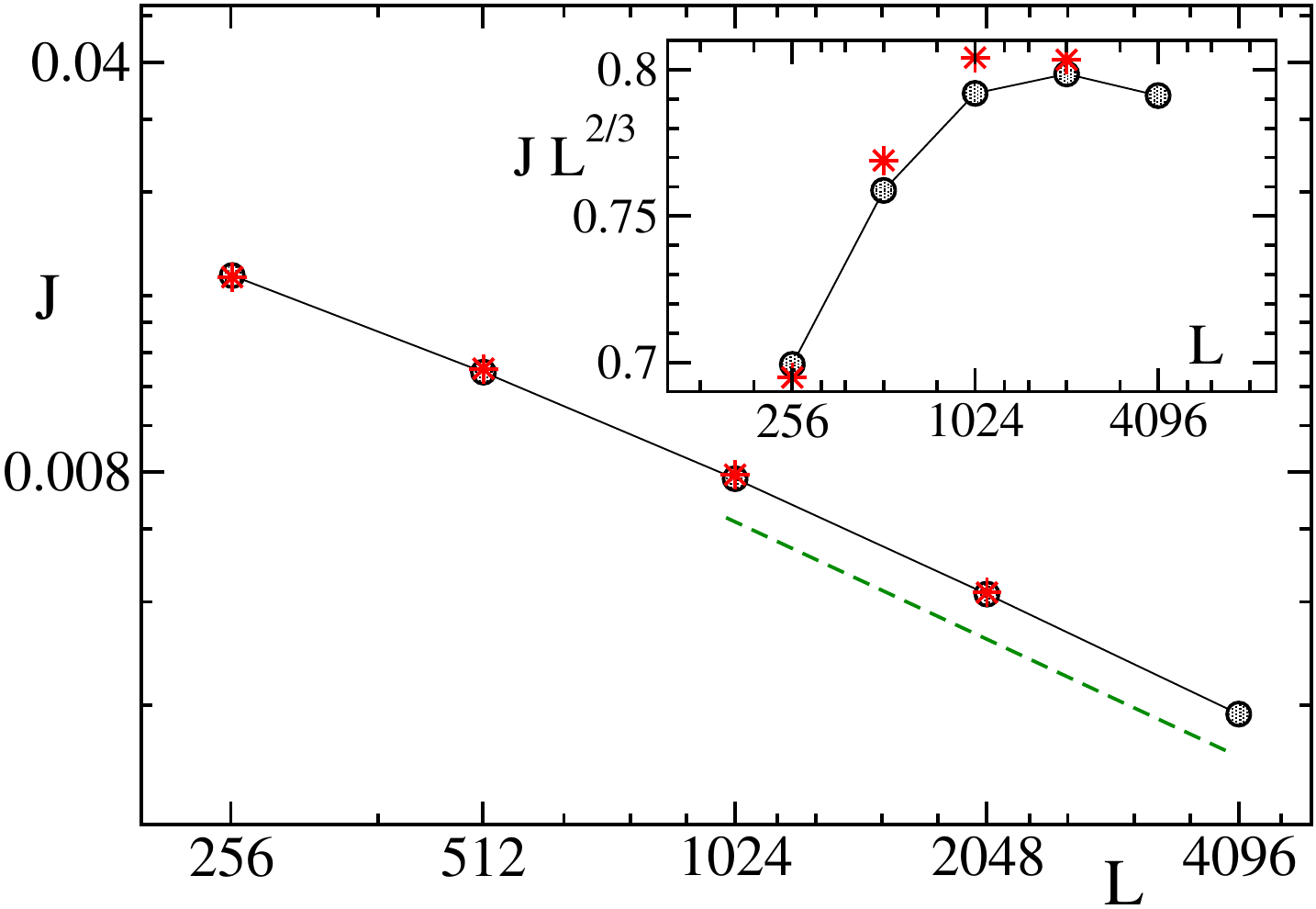}
\caption{Energy flux versus the chain length for the
soft and the hard version of the ding-a-ling type model of Ref.~\cite{Lee-Dadswell2010} 
(in the inset the flux is scaled by a factor $L^{2/3}$).}
\label{fig:canad_flux}
\end{figure}

This expectation is confirmed by the data reported in Fig.~\ref{fig:canad_flux},
where the flux $J$ (black dots) decreases approximately as $L^{-2/3}$  (see the dashed line).
This is more quantitatively confirmed in the inset, where we see that $JL^{2/3}$ saturates and remains constant.
The exponent $2/3$ is the value which corresponds to $\alpha=1/3$:
the characteristic rate of the universality class of asymmetric effective potentials. 

Since these results contradict the claims made in \cite{Lee-Dadswell2010}, I have also investigated a softened version of
the model, where the interactions between free and harmonic particles are ruled by the smooth repulsive potential
(\ref{eq:soft_pot}),
where $C$ has been set equal $10^{-3}$ to make it resemble hard elastic collisions. 
The advantage of this model is that there is no need to accurately identify the collision times;
the disadvantage is that the integration time step must be very small, in order to describe accurately the bouncing process 
(the time step was  $\delta t = 5\ 10^{-4}$).

The outcome of the simulations is reported again in Figs.~\ref{fig:canad},\ref{fig:canad_flux}. 
The temperature profile reported in the first figure (red curve) is 
slightly lower than in the previous case, but the overall shape is very similar. 
As for the energy flux reported in the second figure (red stars), we even observe a semi-quantitative agreement. 
The most important point is,
however, that heat conductivity behaves again in an anomalous way, in agreement with the theoretical expectations.
The motivation for the different results reported in \cite{Lee-Dadswell2010} is unclear; possibly an improper definition
of the local heat flux.
Nevertheless, it is now clear that this ding-a-ling type model exhibits an anomalous, diverging conductivity.

\section{Conclusion}
\label{sec:conclusion}
Fluctuating hydrodynamics shows that three conservation laws (energy, linear momentum, and stretch) are
a prerequisite for the emergence of anomalous heat conductivity in one dimension.
Several publications have anyhow challenged this scenario, providing examples of finite conductivity
in models where a divergence should be instead observed.
I have revisited some of them, showing that in all cases heat conductivity diverges in the thermodynamic limit.
In the case of the ding-a-ling type model studied in Ref.~\cite{Lee-Dadswell2010}, the discrepancy is
due to some kind of error presumably related to the definition of the local flux.
More interesting is the case of non-binding potentials, where I find that the conductivity is
affected by very strong finite-size effects, which mask the asymptotic behavior.
This conclusion is supported by the observation that the effective conductivity can be well approximated
as the sum of two distinct contributions: a kinetic diffusive term and the anomalous one.
This assumption, previously proposed to explain the behavior of nearly integrable systems,
proves, indeed, more general and is valid even in fully chaotic models,

As a result, a question opens up: how can it be that some modes can propagate over very
long distances in the presence of an evident non-integrability? 
In a sense, this unexpected scenario is complentary to the anomalous (non ballistic) divergence observed 
in the symmetric gard-point chain, where the integrals of motions (the {\it velocitons})
diffuse rather than propagating ballistically.
An explanation of when and how this scenario emerges in non-binding potential is a point that
will be worth exploring.

Finally, while representing the conductivity as the sum of two terms,
I have also found that the amplitude $\sigma$ of the hydrodynamic term plays a crucial role
in determining the crossover length.
In one of the non-binding potentials, it is very small, thus contributing to further
increase the crossover length. Understanding which dynamical properties are responsible
for the amplitude of $\sigma$ is another point that should be clarified.

\ack
The author thanks the StatPhys29 organizers, since their invitation has obliged him to revisit
many awkward results which appeared in the last years in the literature.

\section*{References}
\bibliographystyle{iopart-num}

\providecommand{\newblock}{}

\end{document}